\documentclass[aps,prd,showpacs,preprintnumber,nofootinbib,amsmath,amssymb,preprint]{revtex4}

\begin{document}

\title{General first-order mass ladder operators for Klein-Gordon fields}

\author{
Vitor Cardoso$^{1,2}$,
Tsuyoshi Houri$^{3}$,
Masashi Kimura$^{1}$
}
\affiliation{${^1}$ CENTRA, Departamento de F\'{\i}sica, Instituto Superior T\'ecnico, Universidade de Lisboa, Avenida~Rovisco Pais 1, 1049 Lisboa, Portugal}
\affiliation{${^2}$ Perimeter Institute for Theoretical Physics, 31 Caroline Street North
Waterloo, Ontario N2L 2Y5, Canada}
\affiliation{${^3}$ Department of Physics, Kyoto University, Kitashirakawa, Kyoto 606-8502, Japan}

\date{\today}

\pacs{04.50.-h,04.70.Bw}

\begin{abstract}
We study the ladder operator on scalar fields, mapping a solution of the Klein-Gordon equation onto another 
solution with a different mass, when the operator is at most first order in derivatives.
Imposing the commutation relation between 
the d'Alembertian, we obtain the general condition for the ladder operator, 
which contains a non-trivial case which was not discussed 
in the previous work~\cite{Cardoso:2017qmj}.
We also discuss the relation with supersymmetric quantum mechanics.
\end{abstract}

\preprint{KUNS-2692}

\maketitle

\section{Introduction}
Recently, the present authors have developed a formulation for ladder operators of the Klein-Gordon equation (KGE), which map a solution of the massive KGE onto another solution with different mass \cite{Cardoso:2017qmj}. We thus call them mass ladder operators.
The key of the formulation is the mass-dependence of the operators; the commutation relation with the d'Alembertian $\square := g^{\mu\nu}\nabla_\mu\nabla_\nu$ is given by
\begin{align}
[\square, {\cal D}] &= \delta m^2 {\cal D} + 2{\cal Q} (\square - m^2) \,,
\label{generalcommutationrelation}
\end{align}
where ${\cal D}$ is the mass ladder operator, $m^2$ and $\delta m^2$ are constants, and ${\cal Q}$ is a differential operator
\footnote{
If there exists 
a ladder operator ${\cal D}$ which shifts the mass squared of the KGE from $m^2$ to $m^2 + \delta m^2$,
we can show the existence of another ladder operator ${\cal D}^*$ 
which shifts the mass squared of the KGE from $m^2 + \delta m^2$ to $m^2$ (see Appendix~\ref{conjugateoperator}). 
}. Note that this commutation relation acts on 
an arbitrary scalar field.
Since Eq.~\eqref{generalcommutationrelation} is written in the form
\begin{align}
\left(\square - (m^2+\delta m^2)\right) {\cal D} = ({\cal D} + 2{\cal Q})\left(\square - m^2 \right) \,,
\label{equationno2}
\end{align}
one can see that, when ${\cal D}$ acts on a solution $\Phi$ of the massive KGE with mass squared $m^2$, 
i.e., $(\square -m^2)\Phi=0$, ${\cal D}\Phi$ is a solution to 
the massive KGE with mass squared $m^2+\delta m^2$,
i.e.,
\begin{align}
 \left(\square - (m^2+\delta m^2)\right)({\cal D}\Phi) = 0 \,.
\end{align}
As applications, the relation between supersymmetric quantum mechanics,~\footnote{
The relation between the massive KGE and a quantum mechanical system for 
the case of the anti-de Sitter ($AdS$) spacetime was 
discussed in Refs.~\cite{Evnin:2015gma, Evnin:2015wyi, Evnin:2016nsz}.
These references also highlighted that the selection rule for resonance modes, 
which is related to the non-linear instability of $AdS$ spacetime~\cite{Bizon:2011gg},
can be understood in terms of a hidden symmetry.
We find that a sufficient condition for the existence of the resonance modes, i.e.,
that the quantity $\Delta$ (defined in Eq.~(2.7) in~\cite{Evnin:2015gma}) is an integer,
coincides with the condition such that the corresponding massive KGE is connected with 
the massless KGE by the mass ladder operator.
While this might be just a coincidence, it seems to suggest some 
relation between the existence of the resonance modes and the mass ladder operator.
}
and the construction of 
Aretakis's constant~\cite{Aretakis:2011ha, Aretakis:2011hc, Lucietti:2012xr} in an extremal black hole
were also discussed~\cite{Cardoso:2017qmj}.

It is remarkable that such ladder operators are the consequence of conformal symmetry of spacetime. It was assumed in Ref.~\cite{Cardoso:2017qmj} that an $n$-dimensional spacetime admits a closed conformal Killing vector $\zeta^\mu$ being an eigenvector of the Ricci tensor $R_{\mu \nu} \zeta^\nu = (n-1)\chi \zeta_\mu$. 
This assumption is satisfied in a maximally symmetric spacetime.\footnote{For example, there exist $n+1$ closed conformal Killing vectors in an 
$n$-dimensional anti-de Sitter spacetime ($AdS_n$). Using them, one can construct $n+1$ mass ladder operators for the KGE on $AdS_n$.}
Acting the ladder operator on a solution of KGE several times, 
the mass $m^2$ can be shifted to the minimum (or maximum) value $m^2_0$,
satisfying $(n-1)^2/4 \ge m^2_0/\chi \ge n(n-2)/4$.
This implies that from the solutions of KGE with masses in this range, one can construct solutions for other masses, outside of this interval.
A natural question is whether the physical properties of the KGE 
with different masses in this interval are different or not.
While the mass ladder operator defined in~\cite{Cardoso:2017qmj}
cannot connect these two masses, one still needs to consider the possibility that
there exists another ladder operator which does so.
This is one of the motivation to consider the general condition for the mass ladder operator.

One may consider similar ladder operators in Riemannian geometry. In that case, the d'Alembertian is replaced with the Laplacian, and the KGE becomes the eigenvalue equation for the Laplacian,
\begin{align}
(\triangle - \lambda) \Phi = 0 \,,
\end{align} 
where $\triangle$ is the scalar-Laplacian and $\lambda$ is its eigenvalue. In two-dimensional sphere, one can construct the ladder operators for the spherical harmonics $Y_{\ell,m}$, which shift the quantum number $\ell$.
Although the ladder operators were already known~\cite{Higuchi:1986wu,Gelderen:1998}, our procedure highlights that their construction and existence stems from the conformal symmetry of sphere.

In this paper, we consider the inverse problem; we investigate the necessary conditions for a spacetime to admit mass ladder operators of the KGE satisfying the commutation relation \eqref{generalcommutationrelation}. In particular, we focus on first-order operators for simplicity. As shown later, 
if the spacetime admits a closed conformal Killing vector and it is an eigen vector of the Ricci tensor, 
the ladder operator derived from the general condition coincides with that in Ref.~\cite{Cardoso:2017qmj}.
This suggests that the KGEs with different masses in the interval 
$(n-1)^2/4 \ge m^2/\chi \ge n(n-2)/4$ have different physical properties, because 
they cannot be connected with the first order ladder operator,
and this interval for the mass characterizes the KGE in this case.
Also, we show that the general condition is of wider applicability than previously anticipated~\cite{Cardoso:2017qmj}.

This paper is organized as follows. 
In Sec. II, we discuss the condition for the first order mass ladder operator.
In Sec. III, we show the relation with the supersymmetric quantum-mechanical system.
In Sec. IV, The case with symmetry operator is discussed.
Sec. V is devoted to summary and discussion.

\section{Mass ladder operators}
\label{sec:2}
Suppose ${\cal D}$ is a first-order differential operator
on an $n$-dimensional spacetime $(M,g_{\mu \nu})$. Without loss of generality, we write ${\cal D}$ as 
\begin{align}
{\cal D} = \zeta^\mu \nabla_\mu + K \,,
\label{operatorD}
\end{align}
where $\zeta^\mu$ and $K$ are a vector field and a function on $M$, respectively.
The commutation relation with the d'Alembertian is given by
\begin{align}
[\square, {\cal D}]  =
2 (\nabla^\mu \zeta^\nu) \nabla_\mu\nabla_\nu
 + (\square \zeta^\nu +  \zeta_\mu R^{\mu \nu}+ 2\nabla^\nu K) \nabla_\nu
 + (\square K)  \,, \label{eq_15}
\end{align}
where the commutator is assumed as acting on a scalar field.
Since ${\cal D}$ is first order, ${\cal Q}$ in Eq.~\eqref{generalcommutationrelation} is required to be a function $Q$ and thus the right-hand side of Eq.~\eqref{generalcommutationrelation} is calculated as
\begin{align}
 \delta m^2 {\cal D} + 2{\cal Q} (\square - m^2) =
 2Q \square +\delta m^2 \zeta^\nu \nabla_\nu
 +(\delta m^2 K- 2 m^2 Q) \,. \label{eq_16}
\end{align}
Substituting Eqs.~\eqref{eq_15} and \eqref{eq_16} into Eq.~\eqref{generalcommutationrelation}, we obtain the conditions (see Appendix \ref{derivationofladdereq} for the detailed discussion)
\begin{align}
\nabla_{(\mu} \zeta_{\nu)} &= Q g_{\mu \nu},
\label{1steqladder}
\\
\square \zeta_\mu + R_\mu{}^\nu\zeta_\nu + 2\nabla_\mu K &= 
\delta m^2 \zeta_\mu,
\label{2ndeqladder}
\\
\square K &= \delta m^2 K-2  m^2 Q\,.
\label{3rdeqladder}
\end{align}
In what follows, we look into these conditions in detail. First of all, Eq.~\eqref{1steqladder} means that $\zeta^\mu$ is a conformal Killing vector field (CKV), and the function $Q$ is given by
\begin{align}
Q = \frac{1}{n} \nabla_\mu \zeta^\mu,
\label{qckv}
\end{align}
which is called the associated function of $\zeta^\mu$ (see also Appendix \ref{appendix:ckvs} for 
the basic property of CKVs).

If $Q = 0$, $\zeta^\mu$ becomes a Killing vector field (KV). In such a case, since we have 
$\nabla_\mu\zeta^\mu = 0$ and $\square \zeta_\mu + R_\mu{}^\nu\zeta_\nu=0$ 
from the divergence of the Killing equation $\nabla_{(\mu}\zeta_{\nu)}=0$, 
Eq.~\eqref{2ndeqladder} becomes $ 2 \nabla_\mu K = \delta m^2 \zeta_\mu$ and hence $\square K = 0$. 
Substituting this into Eq.~\eqref{3rdeqladder}, we obtain that, if $\delta m^2\neq 0$, then $K=0$. 
Thus, Eq.~\eqref{2ndeqladder} leads $\zeta^\mu = 0$, however, 
this is not the case we are interested in.
We thus find that, if $Q=0$, Eqs.~\eqref{1steqladder}--\eqref{3rdeqladder} 
have the only trivial solution $\delta m^2 = 0$ and $K = {\rm const}$ for a KV $\zeta^\mu \neq 0$. 
This has the well-known consequence that, if $\zeta^\mu$ is a KV, 
${\cal D}=\zeta^\mu\nabla_\mu$ does not shift the mass.
Hereafter, we focus on the non-trivial case when $Q \neq 0$.

The divergence of Eq. \eqref{1steqladder} leads to
\begin{align}
\square \zeta_{\mu} 
+ 
R_{\mu\nu}\zeta^\nu
 &= \left(2 - n\right)\nabla_\mu Q.
\label{eq1steqriccixx}
\end{align}
Substituting Eq.~\eqref{eq1steqriccixx} into Eq.~\eqref{2ndeqladder}, we have
\begin{align}
\delta m^2 \zeta_\mu &= 
\nabla_\mu
\left(
\left(2 - n\right) Q +   2K 
\right) \,.
\label{eq:udivnoshikixx}
\end{align}
Since we are interested in a ladder operator which shifts the mass of KGE, 
we consider the case $\delta m^2 \neq 0$.\footnote{
In the case $\delta m^2=0$, the operator ${\cal D}$ is called a symmetry operator.
While a symmetry operator does not shift the mass, 
it is still interesting since it can map a solution of KGE into another solution of KGE with the same mass.
This case will be discussed in Sec.~\ref{sec:symmoperator}.}
In this case, 
Eq.~\eqref{eq:udivnoshikixx} shows that
$\zeta_\nu$ is closed, i.e., $\nabla_{[\mu} \zeta_{\nu]} = 0$. Hence $\zeta^\mu$ is a closed conformal Killing vector (CCKV) which satisfies the equation
\begin{align}
\nabla_\mu \zeta_\nu = Q g_{\mu \nu}. \label{CCKVequation}
\end{align}
Thus, we find that a spacetime admitting a first-order mass ladder operator of the massive KGE must admit a CCKV.

A CCKV $\zeta^\mu$ with a nonzero associated function $Q$ must be either timelike or spacelike. If $\zeta^\mu$ is null, $0=\nabla_\mu (\zeta^\nu\zeta_\nu) = 2Q \zeta_\mu$ and hence $Q=0$. Since $\zeta_\mu$ is a closed 1-form, one can introduce a function $\lambda$ such that $d\lambda=\zeta_\mu dx^\mu$ in local coordinates $x^\mu$. Using this function $\lambda$ as one of local coordinates, we obtain the local form of metrics admitting a CCKV (see Ref.\ \cite{Batista:2014uja}),
\begin{align}
g_{\mu\nu}dx^\mu dx^\nu 
= \frac{d\lambda^2}{f(\lambda)} + f(\lambda) \tilde{g}_{ij}dx^i dx^j,
\label{standardmetric}
\end{align}
where $f(\lambda)$ is a function depending only on $\lambda$, $\tilde{g}_{ij}$ is an $(n-1)$-dimensional metric\footnote{If $\zeta^\mu$ is timelike, i.e., $f$ takes negative value, 
$(-\tilde{g}_{ij})$ should be positive definite metric
so that the whole metric has $[-,+,\cdots,+]$ signature.}, and $x^\mu=(\lambda,x^i)$ are the local coordinates. In these coordinates, we have $\zeta^\mu (\partial/\partial x^\mu) = f (\partial/\partial \lambda)$ and $Q=f'/2$.
With this local form of the metric, the KGE can be solved by separation of variables between the coordinate $\lambda$ and the others.~\footnote{Moreover, it is shown that if $f'''=0$, the equation separated with  $\lambda$ results in Legendre's differential equation, which is explicitly solved by the Legendre functions (see Appendix \ref{separationofvariableskge}).}

Now that $\zeta_\mu dx^\mu=d\lambda$, Eq.~\eqref{eq:udivnoshikixx} reads $d(\delta m^2 \lambda)=d((2-n)Q+2K)$ where $\delta m^2\neq 0$. Integrating it, we have
\begin{align}
\delta m^2 \lambda = (2-n) Q +2  K  + c,
\label{equation20}
\end{align}
with an integration constant $c$. This means that, since $Q$ is a function of $\lambda$, $K$ is also a function of $\lambda$ and is given by
\begin{align}
 K = \frac{(n-2)}{4} f' 
 + \frac{\delta m^2}{2} \lambda - \frac{c}{2} \,.
\end{align}
Hence, the mass ladder operator is provided in general by
\begin{equation}
 {\cal D} = f \frac{\partial}{\partial\lambda}
 + \frac{(n-2)}{4} f' + \frac{\delta m^2}{2} \lambda - \frac{c}{2} \,.
 \label{def_massladderoperator}
\end{equation}

It is also shown from the divergence of Eq.~\eqref{CCKVequation} that
\begin{align}
\nabla_\mu Q 
 &= \frac{1}{1 - n} R_{\mu\nu} \zeta^\nu \,,
\label{nablaqeq16}
\end{align}
which is same as Eq.~\eqref{eq1steqriccixx} in the case when $\zeta^\mu$ is a CCKV. If we use the metric form \eqref{standardmetric}, it becomes
\begin{align}
R_{\mu\nu}\zeta^\nu = -\frac{(n-1)}{2}f'' \zeta_\mu \,.
\label{eigenvectorfprime}
\end{align}

The remaining condition to solve is Eq.~\eqref{3rdeqladder}.
Since the d'Alembertian acting on $K$ is calculated as
\begin{align}
\square K = f^{-(n-2)/2} \frac{d}{d\lambda} \left(
f^{n/2} \frac{d}{d\lambda} K
\right)
= 
f K^{\prime \prime} + \frac{n}{2}f^\prime K^\prime, 
\end{align}
Eq.~\eqref{3rdeqladder} becomes
\begin{align}
(n-2) f f^{\prime \prime \prime}
+ 
\frac{n(n-2)}{2} f^\prime f^{\prime \prime}
+
\alpha_1 f^\prime
+
\alpha_2 + \alpha_3 \lambda
= 0 \,,
\label{ladderconditionc1c2c3c4}
\end{align}
where
\begin{align}
\alpha_1 = 2( \delta m^2 + 2 m^2)\,, \qquad
\alpha_2 = 2 \delta m^2 c\,, \qquad
\alpha_3 = - 2 (\delta m^2)^2 \,.
\label{alpha123def}
\end{align}
In the next subsections, we explicitly solve this nonlinear differential equation for $f$ by dividing it into two cases: (A) $f^{\prime \prime \prime} = 0$ and (B) $f^{\prime \prime \prime} \neq 0$.

\subsection{$f''' = 0$ case}

If $f^{\prime \prime \prime} = 0$, $f$ is quadratic in $\lambda$. By setting 
\begin{align}
f = c_0 + c_1 \lambda -\chi \lambda^2
\label{fform}
\end{align}
with constants $c_0$, $c_1$ and $\chi$, Eq.~\eqref{ladderconditionc1c2c3c4} becomes
\begin{align}
\alpha_2 + c_1 (\alpha_1 - n(n-2)\chi) + \lambda \Big(
\alpha_3 - 2\chi (\alpha_1 - n(n-2) \chi)
\Big)
=0.
\label{alphadef}
\end{align}
For this equation to be satisfied for any value of $\lambda$, we obtain
\begin{align}
m^2 = - \chi k(k+n-1) \,, \quad
 \delta m^2 = \chi(2k+n-2) \,, \quad
 c = \frac{(2k+n-2)c_1}{2} \,,
\label{massparameterk}
\end{align}
with a parameter $k$. Imposing the condition such that 
both $m^2$ and $m^2 + \delta m^2$ are real values, we find that the parameter $k$ also takes a real value.
Solving the first equation in Eq.~\eqref{massparameterk} w.r.t. $k$,
we obtain 
\begin{align}
k = \frac{1 - n \pm \sqrt{(n-1)^2 - 4m^2/\chi}}{2}.
\end{align}
From the positivity of the inside of the square root of this equation,
we find 
the range of masses so that the ladder operator exists as 
\begin{equation}
\frac{\chi(n-1)^2}{4} \leq m^2 \,, \quad (\chi<0) \,, \qquad
\frac{\chi(n-1)^2}{4} \geq m^2 \,, \quad (\chi>0) \,.
\end{equation}
We note that the lower bound for $\chi < 0$ case is a negative value.
In $AdS$ case, this coincides with the 
Breitenlohner-Freedman bound~\cite{Breitenlohner:1982jf, Breitenlohner:1982bm}
which is the lowest mass for avoiding unstable modes.
To summarize, we have obtained the mass squared $m^2$ and its shift $\delta m^2$ as one-parameter families, which enables us to shift various masses with the mass ladder operator.

In the present case, we can confirm from Eq.~\eqref{eigenvectorfprime} that
$\zeta^\mu$ is an eigenvector of the Ricci tensor,
\begin{align}
R^\mu{}_\nu \zeta^\nu = (n-1)\chi \zeta^\mu \,,
\label{riccieigenvector}
\end{align}
which is the condition assumed in Ref.~\cite{Cardoso:2017qmj}.
Since we also find that $K = - k Q$, the mass ladder operator ${\cal D}$ is given by
\begin{align}
{\cal D} = \zeta^\mu\nabla_\mu - k Q \,,
\label{ladderdk}
\end{align}
which is the mass ladder operator $D_k$ discussed in \cite{Cardoso:2017qmj}.

\subsection{$f''' \neq 0$ case}
Let us consider the case when $f^{\prime \prime \prime} \neq 0$. This case happens only when $n \ge 3$ since
Eq.~\eqref{ladderconditionc1c2c3c4} with $n = 2$ implies $f^{\prime \prime \prime} = 0$. In this case, $f^\prime$, $\lambda$ and a nonzero constant must be linearly independent; otherwise, we have $f'=b_1\lambda+b_2$ with some constants $b_1$ and $b_2$ and hence $f'''=0$. 
We can write Eq.~\eqref{ladderconditionc1c2c3c4} in the form
\begin{align}
-(n-2)\left( f f^{\prime \prime \prime} + \frac{n}{2} f^\prime  f^{\prime \prime} \right) = \alpha_1 f^\prime 
+ \alpha_2 + \alpha_3 \lambda. \label{eq_29}
\end{align}
For a function $f$ fixed, the coefficients $\alpha_1$, $\alpha_2$, and $\alpha_3$ must be determined uniquely because
$f^\prime, \lambda$ and a nonzero constant are linearly independent.
Solving Eq.~\eqref{alpha123def} w.r.t. $m^2, \delta m^2$ and $c$, we obtain
\begin{align}
m^2 = m^2_{\pm} := \frac{\alpha_1 \pm \sqrt{-2\alpha_3}}{4},~~
\delta m^2 = \delta m^2_{\pm} := \mp \frac{\sqrt{-\alpha_3}}{\sqrt{2}},~~
c = c_\pm := \mp \frac{\alpha_2}{\sqrt{-2 \alpha_3}}\,.
\end{align}
This means that for a fixed spacetime, i.e., for a fixed function $f$, 
$m^2$ can take only two values.
{}From Eq.~\eqref{equation20}, the mass ladder operators are given by
\footnote{
Note that 
${\cal D}_-$ is the conjugate ladder operator to $-{\cal D}_+$ (see Appendix~\ref{conjugateoperator}).
}
\begin{align}
{\cal D} = {\cal D}_\pm := \zeta^\mu\nabla_\mu 
+ \frac{(n-2)Q + \delta m^2_\pm \lambda - c_\pm}{2} \,. \label{ladder_fprime3nonzero}
\end{align}
Thus, we obtain mass ladder operators of the massive KGE with only two fixed masses $m^2_{\pm}$
in contrast to the case $f^{\prime \prime \prime} = 0$. 
We can see that ${\cal D}_\pm$ maps a solution of KGE with $m_\pm^2$ to that with $m_\mp^2$ because
we have a relation $m_\pm^2 + \delta m^2_\pm = m_\mp^2$.

If we assume that the function $f$ can be expanded around $\lambda=0$,
\begin{align}
f = \sum_{i = 0}^\infty \beta_i \lambda^i,
\end{align}
Eq.~\eqref{ladderconditionc1c2c3c4} leads to the equations
\begin{align}
-\alpha_2 - \alpha_1\beta_1 -(n-2)(n\beta_1\beta_2 + 6\beta_0\beta_3) 
&= 0 \,, \label{eq_31}
\\
-\alpha_3 - 2 \alpha_1\beta_2 - (n-2)\Big(2n\beta_2^2 + 3(2+n)\beta_1\beta_3 + 24\beta_0\beta_4\Big) &= 0 \,, \label{eq_32}
\\
 \alpha_1  (k-2)\beta_{k-2}
+
\sum_{i=0}^{k} i(i-1) 
\left(
i-2 + \frac{n}{2}(k-i)
\right)\beta_{k-i}  
\beta_{i} 
&= 0 \,. \quad (k \ge 5) \label{eq_33}
\end{align}
There are six constant degrees of freedoms in these equations.
For example, if we fix the constants 
$\alpha_1, \alpha_2, \alpha_3, \beta_0, \beta_1, \beta_2$, 
then $\beta_{k \ge 3}$ will be determined.
We can see that $\beta_{k \ge 4}\neq 0$ if $\beta_3 \neq 0$.
If we assume that $\beta_{k \ge 3} = 0$, the above conditions reduce to the case of $f^{\prime \prime \prime} = 0$. Alternatively, one can interpret them as the equations that determine the values of $m^2$, $\delta m^2$ and $c$. Actually, this can be done with Eqs.~\eqref{eq_31} and \eqref{eq_32} if the values of the six parameters $\beta_0$, $\beta_1$, $\beta_2$, $\beta_3$, $\beta_4$ and $\beta_5$ are provided; the other constants $\beta_i$ $(i\geq 6)$ are determined by Eq.~\eqref{eq_33}. Thus, Eq.~\eqref{eq_33} provides the condition for the metric to admit the mass ladder operator \eqref{ladder_fprime3nonzero}.

While the solution $f$ to Eq.~\eqref{eq_29} had only to be described as a series expansion in a generic case, Eq.~\eqref{eq_29} in the case of $n=6$ and $\alpha_1=0$ can be solved explicitly. Actually, in that case, Eq.~\eqref{eq_29} becomes
\begin{equation}
 (f^2)''' = 4m^4 \lambda + 2m^2c \,,
\end{equation}
where we have used the condition $\alpha_1=0$, i.e., $\delta m^2 = -2m^2$, and this is easily solved by 
\begin{equation}
f = \pm\sqrt{e_0 + e_1\lambda + e_2\lambda^2
 + \frac{m^2c}{3}\lambda^3 + \frac{m^4}{6}\lambda^4} \,, \label{def_h}
\end{equation}
with integration constants $e_0$, $e_1$ and $e_2$.

\section{Relation with supersymmetric quantum mechanics}
\label{sec:susyquantummechanics}

In \cite{Cardoso:2017qmj} a relation of the KGE admitting a mass ladder operator to the Schr\"odinger equation in a supersymmetric quantum-mechanical system was discussed, in which the mass ladder operator is mapped into the supercharge. We repeat this discussion in the present general framework.

For the metric \eqref{standardmetric}, we consider the conformal transformation $\bar{g}_{\mu \nu} = \Omega^2 g_{\mu \nu}$ with the conformal factor $\Omega = 1/\sqrt{f}$. Since the resulting metric is given by
\begin{equation}
 \bar{g}_{\mu \nu} dx^\mu dx^\nu
 = \frac{d\lambda^2}{f(\lambda)^2} + \tilde{g}_{ij}dx^idx^j 
 = d\bar{\lambda}^2 + \tilde{g}_{ij}dx^idx^j \,,
\end{equation}
the CCKV $\zeta^\mu$ on $(M,g_{\mu\nu})$, which is given by $\zeta^\mu \partial_\mu = f \partial_\lambda = \partial_{\bar{\lambda}}$, becomes a KV on $(M,\bar{g}_{\mu \nu})$. Under this conformal transformation the KGE on $g_{\mu\nu}$ is written in the form
\begin{align}
(\square - m^2)\Phi = \Omega^{(n+2)/2} (\bar{\square} - U)\bar{\Phi}\,,
\end{align}
where $\bar{\square}$ is the d'Alembertian for $\bar{g}_{\mu \nu}$, $\bar{\Phi} = \Omega^{(2-n)/2}\Phi$ and
\begin{align}
U = \frac{1}{16}\Big(16 m^2 f + (n-2)^2 (f^\prime)^2 + 4(n-2)f f^{\prime \prime}\Big) \,.
\label{potentialV}
\end{align}
This means that the KGE on $g_{\mu \nu}$, $(\square - m^2)\Phi = 0$ is mapped into the equation on $\bar{g}_{\mu \nu}$,
\begin{align}
(\bar{\square} - U)\bar{\Phi} 
=
f \frac{\partial}{\partial \lambda} \left(
f \frac{\partial}{\partial \lambda}\bar{\Phi} 
\right) 
+
\tilde{\square} \bar{\Phi} 
- U \bar{\Phi}= 0\,,
\label{barkge2}
\end{align} 
where $\tilde{\square}$ is the d'Alembertian for $\tilde{g}_{\mu \nu}$. 
Eq.~\eqref{barkge2} admits separation of variable for the ansatz
\begin{equation}
 \bar{\Phi}(\lambda,x^i)=\psi(\lambda)\Theta(x^i) \,,
 \label{separation_of_variables_2}
\end{equation}
and it then reduces to the ODE
\begin{equation}
- f \frac{d}{d \lambda} \left(
f \frac{d}{d \lambda}\psi  
\right) 
+ U \psi = \nu^2 \psi  \,, \label{barkge3}
\end{equation}
where $\nu^2$ is the separation constant.

Recall that the general condition for the existence of mass ladder operators was given by Eq.~\eqref{ladderconditionc1c2c3c4}. Integrating Eq.~\eqref{ladderconditionc1c2c3c4}, we have
\begin{align}
(n-2) f f^{\prime \prime}
+
\frac{(n-2)^2}{4}(f^\prime)^2 
+ 2(\delta m^2 + 2m^2)f + 2\delta m^2 c \lambda - (\delta m^2)^2\lambda^2
+  \sigma = 0,
\label{integralladderconditionc1c2c3c4}
\end{align}
where $\sigma$ is an integration constant.
Substituting this equation into Eq.~\eqref{potentialV}, we obtain
\begin{align}
U =  \frac{1}{4} \Big(
-2 \delta m^2 f + \delta m^4 \lambda^2 - 2 \delta m^2 c \lambda - \sigma
\Big). \label{originalpotential}
\end{align}
It is remarkable that for this potential, 
Eq.~\eqref{barkge3} is factorized
\footnote{The factorization of Eq.~\eqref{barkge3} can be made for the potential \eqref{originalpotential} with a generic $f$; however, we suppose here that $f$ is a solution to Eq.~\eqref{integralladderconditionc1c2c3c4}.} as
\begin{align}
\left(-f\frac{d}{d\lambda} + W \right)
\left(f\frac{d}{d\lambda} + W \right)\psi 
= E  \psi , \label{barkge4}
\end{align}
where 
\begin{align}
W = \frac{1}{2}\delta m^2 \lambda - \frac{c}{2},
\end{align}
and $E := \nu^2 +(\sigma + c^2)/4$. \footnote{
In terms of the coordinate $\bar{\lambda}$
defined by $d\bar{\lambda} = d\lambda/f$, 
Eq.~\eqref{barkge4} is written as $(-d/d\bar{\lambda} + W)(d/d\bar{\lambda} + W) \psi  = E  \psi $. Then Eq.~\eqref{integralladderconditionc1c2c3c4} is regarded as the condition which determines $W$ as a function of $\bar{\lambda}$.
}
This factorization is important to see a relation with a supersymmetric quantum-mechanical system. We introduce the supercharges
\footnote{
In this section, the dagger $\dag$ denotes the adjoint operator 
in $(M,\bar{g}_{\mu \nu})$ defined in~\cite{Wald:1978vm}.
If we set appropriate Hilbert spaces ${\cal H}_1$ 
and ${\cal H}_2$, $A^\dag$ can be the adjoint operator of $A$ in the usual sense, that is, 
$A$ ($A^\dag$) is a linear map 
from the Hilbert space ${\cal H}_1$ (${\cal H}_2$) to ${\cal H}_2$ (${\cal H}_1$) and 
satisfies the condition $<\!\psi_1,A^\dag \psi_2\!>_1 = <\!A\psi_1,\psi_2\!>_2$,  
where $\psi_i\in {\cal H}_i$ and $<\,,\,>_i$ is the inner product 
in the Hilbert sapce ${\cal H}_i$. In the present case, $<\,,\,>_i$ should be 
defined as $<\!\psi_i,\phi_i\!>_i := \int d\lambda f^{-1} \psi_i \phi_i$ 
where $\psi_i, \phi_i \in {\cal H}_i$.}
\begin{equation}
 A^\dag = -f\frac{d}{d\lambda} + W \,, \qquad
 A = f\frac{d}{d\lambda} + W \,, \label{def_supercharge}
\end{equation}
and set the first Hamiltonian $H$ by $H := A^\dag A$. Then
\begin{equation}
 H = -f\frac{d}{d\lambda}\left(f\frac{d}{d\lambda}\right) + V \,,
\end{equation}
with 
\begin{align}
V :=  \frac{1}{4} \Big(
-2 \delta m^2 f + \delta m^4 \lambda^2 - 2 \delta m^2 c \lambda + c^2
\Big).
\end{align}
Eq.~\eqref{barkge4} is written as
\begin{equation}
 H \psi = E \psi \,. \label{H1_equation}
\end{equation}
Setting the second Hamiltonian $\tilde{H}$ by $\tilde{H} := A A^\dag$, we have
\begin{equation}
 H = -f\frac{d}{d\lambda}\left(f\frac{d}{d\lambda}\right) + \tilde{V} \,,
\end{equation}
where
\begin{align}
\tilde{V} := \frac{1}{4} \Big(
2 \delta m^2 f + \delta m^4 \lambda^2 - 2 \delta m^2 c \lambda + c^2
\Big).
\end{align}
We consider the eigenvalue equation for $\tilde{H}$
\begin{equation}
 \tilde{H} \tilde{\psi} = \tilde{E} \tilde{\psi} \,. \label{H2_equation}
\end{equation}

Now, by construction, we have the so-called intertwining relations
\begin{equation}
  \tilde{H} A = A H \,, \qquad
  H A^\dag = A^\dag \tilde{H} \,,
\end{equation}
which shows that if $\psi$ ($\tilde{\psi}$) is an eigenfunction for $H$ ($\tilde{H}$), $A\psi$ 
($A^\dag \tilde{\psi}$) is an eigenfunction for $\tilde{H}$ ($H$) with the same eigenvalue. 
Indeed, we can easily check that
\begin{eqnarray}
&& \tilde{H} A\psi = A H \psi = A(E \psi) 
 =  E (A\psi) \,, \\
&& H A^\dag \tilde{\psi} = A^\dag \tilde{H} \tilde{\psi} = A^\dag( \tilde{E} \tilde{\psi}) 
 =  \tilde{E} (A^\dag \tilde{\psi}) \,.
\label{eq57h2}
\end{eqnarray}
If $f$ is a solution to Eq.~\eqref{integralladderconditionc1c2c3c4} with the set of parameters $(m^2, \delta m^2, c)$, 
it is also a solution to Eq.~\eqref{integralladderconditionc1c2c3c4} with $(m^2 + \delta m^2, -\delta m^2, -c)$. 
This means that, for a spacetime fixed (i.e., a solution $f$ fixed), 
we have ladder operators for the KGE with mass squared $m^2$ and $m^2+\delta m^2$, 
which map solutions with the mass squared $m^2$ and $m^2+\delta m^2$ into solutions 
with $m^2+\delta m^2$ and $m^2$, respectively. 
As already seen the KGE with mass squared $m^2$ became Eq.~\eqref{H1_equation}; 
whereas, since we have $A^\dag \to -A$ and $A\to -A^\dag$ under 
the transformation $(m^2, \delta m^2, c) \to (m^2 + \delta m^2, -\delta m^2, -c)$, 
$H \to \tilde{H}$, and $E \to E$ under this transformation,
and hence the KGE with mass squared $m^2+\delta m^2$ becomes Eq.~\eqref{H2_equation}
with $\tilde{E} = E$.
Thus, we see that the supercharges mapping eigenfunctions between two Hamiltonians $H$ and $\tilde{H}$ 
correspond to the mass ladder operators mapping solutions between two KGEs with mass squared $m^2$ and
$m^2 + \delta m^2$.\footnote{
We can also show this by an explicit calculation.
We adapt the notation like $H^{(m^2,\delta m^2, c)}$
which denotes an operator when 
the mass ladder operator exists for the KGE with mass squared $m^2$
and has parameters $\delta m^2$ and $c$.
Here, we only consider a fixed function $f$ and a fixed value of $\sigma$.
From the above discussion, there also exists $H^{(m^2 + \delta m^2,-\delta m^2, -c)}$,
and we can see the relation $H^{(m^2 + \delta m^2,-\delta m^2, -c)} = \tilde{H}^{(m^2,\delta m^2,c)}$. 
For a function $\psi_{m^2}$ 
which satisfies $H^{(m^2,\delta m^2, c)}\psi_{m^2} = E^{(m^2,\delta m^2, c)}\psi_{m^2}$,
we have
\begin{align}
H^{(m^2 + \delta m^2,-\delta m^2, -c)} 
A^{(m^2,\delta m^2,c)}\psi_{m^2}
&= 
\tilde{H}^{(m^2,\delta m^2,c)} A^{(m^2,\delta m^2,c)}\psi_{m^2}  
\notag \\&=
A^{(m^2,\delta m^2,c)} H^{(m^2,\delta m^2,c)} \psi_{m^2} 
\notag \\&= 
E^{(m^2,\delta m^2,c)}A^{(m^2,\delta m^2,c)}\psi_{m^2}
\notag \\&= 
E^{(m^2 + \delta m^2,-\delta m^2, -c)}A^{(m^2,\delta m^2,c)}\psi_{m^2},
\end{align}
where 
we used $E^{(m^2,\delta m^2,c)} = E^{(m^2 + \delta m^2,-\delta m^2, -c)}$
since it only depends on $c^2$.
This shows that $\psi_{m^2}$ is 
mapped into a solution of KGE with mass squared $m^2 + \delta m^2$ by $A^{(m^2,\delta m^2,c)}$.
}
More explicitly, we can show the relation between the mass ladder operator and the supercharge. The mass ladder operator is given by \eqref{def_massladderoperator}, and the supercharge $A$ is given by \eqref{def_supercharge}. When we compare both expressions, we find
\begin{equation}
 A \psi = {\cal D} \Xi \,,
\end{equation}
where $\Xi =\Omega^{(n-2)/2}\psi$ is the function of $\lambda$ separated in a solution $\Phi$ to the KGE as \eqref{separation_of_variables} (see Appendix C for details).

Finally we remark that, in the $f''' = 0$ case, two Hamiltonians in the supersymmetric quantum mechanics 
admit shape invariance, so that the supercharges come to map wave functions to wave functions for 
the same Hamiltonian but between different energies (see Ref.\ \cite{Cardoso:2017qmj} and 
see also Appendix~\ref{appendix:shapeinvariance}). Due to this property we can construct 
all the energy spectrum for the system from a seed wave function by using the supercharge. 
This corresponds to shifting the masses of Klein-Gordon fields by acting the mass ladder operator 
many times. On the other hand, in the $f'''\neq 0$ case, the present system does not admit shape invariance. 
In that case, as was seen above, since the supercharges only map wave functions between 
two Hamiltonian $H$ and $\tilde{H}$ in the same energy level, the mass ladder operators also 
only map solutions between two KGEs with mass squared $m^2$ and $m^2+\delta m^2$.

\section{Symmetry operators}
\label{sec:symmoperator}
We now specialize our discussion to the $\delta m^2 = 0$ case. 
The discussion is identical to the previous one, until Eq.~\eqref{eq:udivnoshikixx}. 
We should note that $\zeta^\mu$ is a CKV, but it is not necessarily closed, since Eq.~\eqref{eq:udivnoshikixx} 
with $\delta m^2 = 0$ does not imply the closed condition for $\zeta^\mu$.
Since we have $\delta m^2=0$, the commutation relation \eqref{generalcommutationrelation} is given by
\begin{equation}
 [\Box,{\cal D}] = 2Q(\Box - m^2) \,.
\end{equation}
In other words, ${\cal D}$ is a symmetry operator of the KGE. 
When ${\cal D}$ acts on a solution $\Phi$ to the KGE with mass squared $m^2$, i.e., $(\square -m^2)\Phi=0$, 
one finds that $(\square - m^2)({\cal D}\Phi) = 0$.

It is shown from Eq.~\eqref{eq:udivnoshikixx} with $\delta m^2 = 0$ that
\begin{align}
2 K + (2 - n) Q  + c = 0,
\label{symmetryeq13}
\end{align}
where $c$ is an integration constant. This constant degree of freedom
corresponds to constant shift symmetry for $K$ when ${\cal D}$ is a symmetry operator $\delta m^2 =0$. 
We can see this from that 
Eqs.~\eqref{2ndeqladder} and \eqref{3rdeqladder} with $\delta m^2 = 0$ only contain derivative of $K$.
So, it is enough to consider the case $c = 0$.
Note that Eq.~\eqref{symmetryeq13} seems to be a particular case of Eq.~\eqref{equation20}, but $\zeta^\mu$ is not necessarily closed now. From Eqs.~\eqref{3rdeqladder} and \eqref{symmetryeq13}, we obtain
\begin{align}
(n-2) \Box Q + 4 m^2 Q = 0 \,. \label{eq_37}
\end{align}
Thus, we find that if there exists a conformal Killing vector with the associated function which 
satisfies Eq.~\eqref{eq_37},
%
\begin{equation}
{\cal D} = \zeta^\mu\nabla_\mu + \frac{n-2}{2}Q\,.
\label{dforsymmetryoperator}
\end{equation}
becomes a symmetry operator of the KGE with mass squared $m^2$. 

We consider several special cases as follows:
\subsection{Killing vector case}
When $\zeta^\mu$ is a KV, 
it satisfies the Killing equation $\nabla_{(\mu}\zeta_{\nu)} = 0$, which corresponds to the $Q=0$ case 
in Sec.~\ref{sec:2}. In this case, we obtain $K={\rm const}$ from Eq.~\eqref{symmetryeq13}.
Since Eq.~\eqref{eq_37} holds for any $m^2$, 
${\cal D}$ becomes a symmetry operator of the KGE with arbitrary mass squared. This is a well-known result.
In this case, the commutation relation is given by $[\Box,{\cal D}]=0$.

\subsection{Homothetic vector case}
When $\zeta^\mu$ is a homothetic vector (HV), 
it satisfies the conformal Killing equation $\nabla_{(\mu}\zeta_{\nu)} = Q g_{\mu\nu}$ with $Q$ constant. 
This case requires that $m^2=0$ and $K={\rm const}$ from Eqs.~\eqref{symmetryeq13} and \eqref{eq_37}. 
So, ${\cal D}$ becomes a symmetry operator of the massless KGE. 
In this case, the commutation relation is given by $[\Box, {\cal D}]=2Q\Box$.

\subsection{Closed and eigenvector of Ricci tensor case}
If we consider that $\zeta^\mu$ is closed and an eigen vector of 
Ricci tensor $R_{\mu \nu}\zeta^\nu = \chi (n-1)\zeta_{\mu}$, 
then we have $\square Q  + \chi n Q = 0$.
Compared with Eq.~\eqref{eq_37}, 
we can see that ${\cal D}$ is a symmetry operator for KGE
with $m^2 = n (n-2)\chi/4$. 
In this case, the ladder operator corresponds to $D_k$ with $k = -(n-2)/2$, 
i.e., $\delta m^2 = \chi (2k + n-2)= 0$ in~\cite{Cardoso:2017qmj}.

\subsection{Constant scalar curvature case}
Using Eq.~\eqref{boxq_ricci}, 
Eq.~\eqref{eq_37} can be written as
\begin{equation}
   {\cal L}_\zeta R = \frac{8 m^2(n-1)- 2(n-2) R}{n-2} Q \,.
\end{equation}
This equation implies that if a spacetime has a constant scalar curvature 
$R = \chi n (n-1)$,
${\cal D}$ becomes a symmetry operator of the KGE with mass squared $m^2 = n (n-2)\chi/4$.
In $n\ge 3$, it is known that all (non-trivial) CKVs are closed in a maximally symmetric spacetime,
so in fact, this symmetry operator coincides with that constructed from CCKV in the previous subsection 
in maximally symmetric case.

\subsection{Two dimensional case}
From Eq.~\eqref{eq_37}, 
we have $m^2=0$ for a CKV with $Q \neq 0$ if $n = 2$.
Thus, the operator ${\cal D} = \zeta^\mu \nabla_\mu$ 
in Eq.~\eqref{dforsymmetryoperator}
becomes a symmetry operator of the massless KGE for any CKV in two dimensions.

\section{Summary and discussion}
In this paper we have investigated mass ladder operators of the massive KGE. In particular, we focused on first-order mass ladder operators which satisfy the commutation relation \eqref{generalcommutationrelation}. We found that if a spacetime admits such a ladder operator, the metric must admit a CCKV and hence the metric is written in the form \eqref{standardmetric}. Moreover, if we assume that the mass ladder operator can act on scalar fields with 
continuous mass range for a fixed back ground spacetime, 
we can say $f^{\prime \prime \prime} = 0$ and obtain the same condition for the mass ladder operator in 
the previous paper~\cite{Cardoso:2017qmj}.
We also obtain the $f^{\prime \prime \prime} \neq 0$ case  
which is not discussed in~\cite{Cardoso:2017qmj}.
In that case, the ladder operator still exists but 
we can construct it for the KGE with two fixed masses in contrast to $f^{\prime \prime \prime} = 0$ case.
We derived the general metric forms for both cases.

The meaning of the conditions for the mass ladder operator
in the previous paper \cite{Cardoso:2017qmj} are now clearer.
As shown in Appendix~\ref{separationofvariableskge}, the presence of
CCKV implies that the KGE has a solution with the form of the separation of variable. 
The additional condition $R_{\mu \nu}\zeta^\nu = (n-1)\chi \zeta_\mu$ 
in Eq.~\eqref{riccieigenvector} (or $f^{\prime \prime \prime} = 0$)
is the condition that the KGE can be solved by using the Legendre function.
In that case, we can 
obtain the mass ladder operator from the differential recursion relations for the Legendre functions.
This is the mathematical reason for the existence of the mass ladder operator in~\cite{Cardoso:2017qmj}.

An interesting result we found concerns the existence condition for the general first-order mass ladder operator:
the corresponding quantum-mechanical system becomes supersymmetric.
The two fixed masses in $f^{\prime \prime \prime} \neq 0$ case correspond to
the superpartners in this quantum-mechanical system.
If the potential also has a shift symmetry, which corresponds to $f^{\prime \prime \prime} = 0$ case,
the mass ladder operator can act on scalar fields with various masses.

As a prospect, there are several possible generalizations of the present work. 
It might be interesting to consider the ladder operator for a system other than the KGE.
As a simple example, we show a mass ladder operator for a scalar field which is coupled with a gauge field 
in Appendix.\ref{generalization:gaugefield}.
In \cite{Higuchi:1986wu} the symmetric tensor harmonics in $S^n$ were constructed by embedding 
$S^n$ into $S^{n+1}$ and using isometry of $S^{n+1}$, which seems to suggest the existence of 
ladder operators between tensor harmonics in a maximally symmetric space. This may be extended 
to the case of the Lorentz signature. Thus, it is interesting to see how far our formulation 
can be applied to vector, tensor and spinor fields. Another generalization is the extension 
to the operator which contain higher derivative. In that case, the general conditions for the 
existence of such an operator will be more complicated; on the other hand, the general conditions 
will contain the Riemann curvature tensor, in contrast to the first-order case 
[cf.~\eqref{1steqladder}--\eqref{3rdeqladder}], which will restrict the space(-time) 
admitting the ladder operator tightly. A naive expectation is that higher-order mass ladder 
operators are related to higher-rank conformal Killing tensors. In a maximally symmetric spacetime, 
any conformal Killing tensor is reducible, that is, decomposable as the symmetric tensor product 
of conformal Killing vector fields. Hence, it is expected that in a maximally symmetric spacetime, 
higher-order ladder operators are also decomposable as the multiple of first order mass ladder operators. 
It is an interesting question to ask whether or not there exists a spacetime admitting an irreducible 
higher-order ladder operator. In any case, such generalizations are challenging, because
the conditions from the commutation relation contains the non-trivial case $f^{\prime \prime \prime} \neq 0$ 
even in the simplest case where the mass ladder operator on scalar contains at most first order derivative. 
It would be important to have an physical insight into them, while calculating the general conditions with power.

\section*{Acknowledgments}
V.C. and M.K. acknowledge financial support provided under the European Union's H2020 ERC 
Consolidator Grant ``Matter and strong-field gravity: New frontiers in Einstein's theory''
grant agreement no. MaGRaTh-646597, and under the H2020-MSCA-RISE-2015 Grant No. StronGrHEP-690904.
T.H. was supported by the Supporting Program for Interaction-based Initiative Team Studies (SPIRITS) 
from Kyoto University and by a JSPS Grant-in-Aid for Scientific Research (C) No.\,15K05051.

\appendix

\section{Conjugate ladder operator}
\label{conjugateoperator}

According to~\cite{Wald:1978vm}, in a spacetime $(M,g_{\mu \nu})$,
for a differential operator ${\cal L}$ which acts on scalar fields, 
we define the adjoint operator ${\cal L}^\dag$ as
\begin{align}
({\cal L}^\dag \phi_2) \phi_1 - \phi_2 {\cal L} \phi_1 = \nabla_\mu s^\mu,
\end{align}
where $\phi_1, \phi_2$ are arbitrary scalar fields, and $s^\mu$ is a vector field.
From this definition, we can show $({\cal L}_1 {\cal L}_2)^\dag =  {\cal L}_2^\dag  {\cal L}_1^\dag$.
We should note that the adjoint operator is defined locally in this definition.

Since we obtain by taking the adjoint of Eq.~\eqref{equationno2} that 
\begin{align}
{\cal D}^\dag \left(\square - (m^2+\delta m^2)\right)
=\left(\square - m^2 \right) ({\cal D} + 2{\cal Q})^{\dag},
\end{align} 
where we have used the fact that $\square$ is self-adjoint, $({\cal D} + 2{\cal Q})^{\dag}$ rather than ${\cal D}^\dag$ is the operator which maps a solution to the massive KGE with mass squared $m^2+\delta m^2$ to that with $m^2$. In this paper, 
\begin{align}
{\cal D}^* := ({\cal D}+2 {\cal Q})^\dag
\end{align} 
will be said to be conjugate to ${\cal D}$. Note that the existence of such 
a conjugate operator comes from the self-adjointness of the d'Alembertian $\square$,
and this discussion holds even if ${\cal D}$ contains higher derivative.
We can see that ${\cal D}^* {\cal D}$ becomes a symmetry operator for the KGE.
When we consider the first-order mass ladder operator, 
${\cal D}$ is given by \eqref{operatorD} with Eqs.~\eqref{1steqladder}--\eqref{3rdeqladder},
and its conjugate operator is explicitly given by 
\begin{align}
{\cal D}^* &= -\zeta^\mu \nabla_\mu + K - (n-2)Q.
\end{align}

\section{Derivation of Eqs.~\eqref{1steqladder}-\eqref{3rdeqladder}}
\label{derivationofladdereq}
First, we introduce the following proposition.
\\\\
{\bf Proposition 1.}~~{\it 
Let $A^{\mu \nu}, B^\mu, C$ be tensor, vector and scalar fields, respectively,
and assume that the metric $g_{\mu \nu}$ is $C^2$.
If the equation 
\begin{align}
A^{\mu \nu} \nabla_\mu \nabla_\nu  \Phi +B^\mu \nabla_\mu \Phi  +C \Phi = 0
\end{align}
holds for any scalar field $\Phi$, the equations $A^{\mu \nu} = B^\mu = C = 0$ hold.
}
\\
{\bf Proof}.~~
By taking the Riemann normal coordinate around a point $p$ on a spacetime, 
the metric behaves
$g_{\mu \nu} = \eta_{\mu \nu} + {\cal O}(x^2)$,
where $\eta_{\mu \nu}$ is a flat metric and we assumed that the point $p$ is the origin of this coordinate.
Choosing a scalar field as $\Phi = c^A_{\mu \nu}x^\mu x^\nu/2 + c^B_{\mu} x^\mu + c^C$
with constants $c^A_{\mu \nu}, c^B_{\mu}, c^C$,
we obtain an equation at $p$ as
\begin{align}
A^{\mu \nu}|_p c^A_{\mu \nu} + B^\mu|_p c^B_{\mu}  + C|_p c^C = 0.
\end{align}
Since $\Phi$ is an arbitrary scalar, this equation holds for any 
$c^A_{\mu \nu}, c^B_{\mu}, c^C$, then 
$A^{\mu \nu}|_p = B^\mu|_p = C|_p = 0.$
Since $p$ is any point on the spacetime, we have
$A^{\mu \nu} = B^\mu = C = 0.$
\hfill$\Box$
\\

Let us consider the action of Eq.~\eqref{generalcommutationrelation} on 
an arbitrary scalar field $\Phi$,
\begin{align}
[\square, {\cal D}] \Phi = (\delta m^2 {\cal D} + 2{\cal Q} (\square - m^2) )\Phi.
\end{align}
Substituting Eqs.~\eqref{eq_15} and \eqref{eq_16} into this equation,
we obtain 
\begin{align}
(\nabla_{(\mu} \zeta_{\nu)} - Q g_{\mu \nu}) \nabla^\mu \nabla^\nu \Phi
+
(\square \zeta_\mu + R_\mu{}^\nu\zeta_\nu + 2\nabla_\mu K 
-
\delta m^2 \zeta_\mu) \nabla_\mu \Phi
&
\notag \\
+
(\square K - \delta m^2 K-2  m^2 Q) \Phi & = 0.
\end{align}
Thus, we obtain Eqs.~\eqref{1steqladder}--\eqref{3rdeqladder} from the above proposition.

\section{Some formulas on conformal Killing vectors}
\label{appendix:ckvs}

It is known that in $n\geq 3$ dimensions, the CKV equation 
$\nabla_{(\mu}\zeta_{\nu)} = Q g_{\mu\nu}$ leads to
\begin{eqnarray}
 \nabla_\mu \zeta_\nu 
&=& L_{\mu\nu} + Q g_{\mu\nu} \,, \label{CKV_def1} \\
 \nabla_\mu L_{\nu\rho}
&=& -R_{\nu\rho\mu}{}^\sigma\zeta_\sigma - 2 g_{\mu[\nu}\eta_{\rho]} \,, \label{CKV_def2} \\
 \nabla_\mu Q
&=& \eta_\mu \,, \label{CKV_def3}\\
 \nabla_\mu \eta_\nu
&=& -(\nabla_\rho S_{\mu\nu})\zeta^\rho - 2S_{\mu\nu}Q
-\frac{2}{n-2} R^\rho{}_{(\mu} L_{\nu)\rho} \,, \label{CKV_def4}
\end{eqnarray}
where $L_{\mu\nu} := \nabla_{[\mu}\zeta_{\nu]}$, $\eta_\mu := \partial_\mu Q$ and 
\begin{equation}
S_{\mu\nu} = \frac{1}{n-2}\left(R_{\mu\nu}-\frac{R}{2(n-1)}g_{\mu\nu}\right)
\end{equation}
is the Schouten tensor with the trace $S=(1/2(n-1))R$. These equations are sometimes called 
the prolonged equations, or, structure equations, which show that a CKV $\zeta^\mu$ is completely 
determined by the values of $\zeta^\mu$, $L_{\mu\nu }$, $Q$ and $\eta_\mu$ at a point on $M$ 
and hence the space of CKVs is a vector space of finite dimensions $(n+1)(n+2)/2$. 
In contrast, the space of CKVs in two dimensions is a vector space of infinite dimensions. 
Actually, since the Riemann tensor is given by 
$R_{\mu\nu\rho\sigma}=(R/2)(g_{\mu\rho}g_{\nu\sigma}-g_{\mu\sigma}g_{\nu\rho})$ 
in two dimensions, Eq.\ (\ref{CKV_def2}) becomes
\begin{eqnarray}
 \nabla_\mu L_{\nu\rho}
&=& - g_{\mu[\nu}\zeta_{\rho]}R - 2 g_{\mu[\nu}\eta_{\rho]} \,,
\end{eqnarray}
and hence we do not have the counterpart to Eq.\ (\ref{CKV_def4}) in two dimensions.

From Eqs.~\eqref{CKV_def1}--\eqref{CKV_def4}, we can derive useful relations
\begin{align}
\nabla_\mu Q &= \frac{1}{n}\nabla_\mu \nabla_\nu \zeta^\nu
=
-\frac{1}{n-2} \left( \square \zeta_\mu + R_{\mu \nu}\zeta^\nu\right),
\\
\square Q &= -\frac{1}{n-1} \left(
\frac{1}{2}{\cal L}_{\zeta} R
+
R Q
\right),
\label{boxq_ricci}
\\
[\nabla_\nu, \square]\zeta^\nu &= \nabla^\mu(R_{\mu \nu}\zeta^\nu).
\end{align}
Moreover, if we impose the closed condition $\nabla_{[\mu}\zeta_{\nu]} = 0$, we have $L_{\mu\nu}=0$. Hence, Eq.~\eqref{CKV_def2} leads to Eq.~\eqref{nablaqeq16}.

\section{Separation of variables in the Klein-Gordon equation}
\label{separationofvariableskge}
The existence of a CCKV restricts the metric form into \eqref{standardmetric}, which leads to the separation of the variable $\lambda$ in the KGE. Actually, since the d'Alembertian for the metric form \eqref{standardmetric} acts on a scalar field $\Phi$ as
\begin{equation}
 \square\Phi = f^{-(n-2)/2}\partial_\lambda (f^{n/2} \partial_\lambda \Phi)
           + f^{-1} \tilde{\square} \Phi \,,
           \label{General_KG_Equation}
 \end{equation}
where $\tilde{\square} := \tilde{g}^{ij}\tilde{\nabla}_i \tilde{\nabla}_j$ and $\tilde{\nabla}_i$ 
are the d'Alembertian and the covariant derivative of $\tilde{g}_{ij}$, the KGE $(\square -m^2)\Phi = 0$ 
admits the separation of variable for the ansatz
\begin{equation}
 \Phi(\lambda,x^i)=\Xi(\lambda)\Theta(x^i) \,,
 \label{separation_of_variables}
\end{equation}
and reduces to the ODE
\begin{equation}
 f^2 \Xi^{\prime\prime} + \frac{n}{2}f f^\prime \Xi^\prime
 +\left(-m^2 f + \nu^2\right)\Xi = 0 \,,
 \label{Laplacian_General_ODE}
\end{equation}
where $\nu^2$ is the separation constant, satisfying the eigenvalue equation 
$\tilde{\square} \Theta = \nu^2 \Theta$. Below, we discuss Eq.~\eqref{Laplacian_General_ODE} 
in the respective case of $f'''=0$ and $f'''\neq 0$. For both cases, the ladder operator 
${\cal D}$ maps a solution $\Xi$ to the equation \eqref{Laplacian_General_ODE} into a solution 
$\overline{\Xi} := {\cal D}\Xi$ to the equation
\begin{equation}
 f^2 \overline{\Xi}^{\prime\prime} + \frac{n}{2}f f^\prime \overline{\Xi}^\prime
 +\left(-(m^2+\delta m^2) f + \nu^2\right)\overline{\Xi} = 0 \,.
 \label{Laplacian_General_ODE_2}
\end{equation}
The difference between two cases is that, in the $f'''=0$ case, $f$ is given independent of 
$m^2$, whereas, in the $f'''\neq 0$ case, $f$ depends on $m^2$.

In the $f^{\prime \prime \prime} = 0$ case, the function $f$ is given by the quadratic 
function (\ref{fform}). Hence, Eq.\ (\ref{Laplacian_General_ODE}) becomes
\begin{eqnarray}
\left(c_0+c_1\lambda -\chi\lambda^2\right)^2 \Xi^{\prime\prime} 
 + \frac{n}{2}\left(c_0+c_1\lambda -\chi\lambda^2\right)
 (c_1-2\chi\lambda) \Xi^\prime && \nonumber\\
 +\left(-m^2 \left(c_0+c_1\lambda -\chi\lambda^2\right) + \nu^2\right)\Xi 
 &=& 0 \,. \label{AppEq_C4}
\end{eqnarray}
This contains 6 parameters; $c_0$, $c_1$, $\chi$, $m^2$, $\nu^2$ and $n$. 
It is interesting that this second-order ODE can be solved by the Legendre functions. 
In fact, we perform the redefinition of the variable $\Xi(\lambda)$ by 
$\Xi(\lambda) = f(\lambda)^{(2-n)/4} P(\lambda)$. 
Moreover, after making the coordinate transformation $\lambda \to z$
\begin{equation}
 \lambda = \frac{c_1-z\sqrt{c_1^2+4c_0\chi}}{2\chi} \,,
\end{equation}
with the reparametrization $(\chi, m^2, \nu^2) \to (\kappa^2, \alpha, \beta)$
\begin{eqnarray}
 \chi &=& \frac{\kappa^2 -c_1^2}{4c_0} \,, \nonumber \\
 m^2  &=& -\frac{(c_1^2-\kappa^2 )(-n(2-n)-4\alpha(1+\alpha))}{16c_0} \,, \nonumber\\
 \nu^2  &=& \frac{\kappa^2 (n-2-2\beta)(n-2+2\beta)}{16} \,, \label{reparam_chimunu}
\end{eqnarray}
Eq.~\eqref{AppEq_C4} reduces to the associated Legendre differential equation
\begin{equation}
  (1-z^2)\frac{d^2P}{dz^2}-2z\frac{dP}{dz}
  +\left\{\alpha(\alpha+1)-\frac{\beta^2}{1-z^2}\right\}P=0 \,,
\end{equation}
which is solved by the Legendre function $P=P_\alpha^\beta(z)$.
Hence, the solution to the equation \eqref{AppEq_C4} is given by
\begin{equation}
 \Xi_{m^2}^{\nu^2}(\lambda)
= C_{m^2, \nu^2} f(\lambda)^{(2-n)/4}P_\alpha^\beta(z(\lambda)) \,, 
\label{Sol_Laplace_General}
\end{equation}
where $z(\lambda)=(c_1-2\chi\lambda)/\sqrt{c_1^2+4c_0\chi}$ and $C_{m^2, \nu^2}$ 
is the normalization constant. Note that $\alpha$ and $\beta$, given by (\ref{reparam_chimunu}), 
are functions of $m^2$ and $\nu^2$, respectively.

The ladder operator in this case is given by
\begin{equation}
 {\cal D} = D_k = f \frac{d}{d\lambda} - \frac{k(c_1-2\chi \lambda)}{2} \,.
\end{equation}
and leads to the relation
\begin{align}
D_k \Xi_{m^2}^{\nu^2}(\lambda,x^i) &= 
\frac{C_{m^2, \nu^2}}{C_{m^2+\delta m^2, \nu^2}}
\Xi_{m^2 + \delta m^2}^{\nu^2}(\lambda,x^i),
\end{align}
where $\Xi_{m^2 + \delta m^2}^{\nu^2}(\lambda,x^i) = 
f(\lambda)^{(2-n)/4}P_{\alpha(m^2 + \delta m^2)}^{\beta(\nu^2)}(z(\lambda))$, and $k$ is chosen to satisfy the relations \eqref{massparameterk}.
This relation is essentially the same result as the well-known differential recursion relations for the Legendre functions.

In $f''' \neq 0$ case, 
Eq.~\eqref{Laplacian_General_ODE} deals with a wider class of second-order ODEs than 
the associated Legendre differential equation.
In this case, the masses we can shift are only two kinds, for a spacetime fixed.

\section{Supersymmetric quantum-mechanical systems with a shape invariance}
\label{appendix:shapeinvariance}

If the function $f$ is given by a quadratic function \eqref{fform}, 
$m^2, \delta m^2, c$ takes Eq.~\eqref{massparameterk},
\begin{align}
m^2 = m^2_k =  - \chi k(k+n-1) \,, \quad
 \delta m^2 = \delta m^2_k = \chi(2k+n-2) \,, \quad
 c  = c_k = \frac{(2k+n-2)c_1}{2} \,,
\notag
\end{align}
where we added the subscript $k$ for the convenience.
{}From Eq.~\eqref{integralladderconditionc1c2c3c4}, $\sigma$ takes
\begin{align}
\sigma = \sigma_k = -\frac{c_1^2 (n-2)^2}{4} + 4 c_0 k (k+n-2)\chi.
\end{align}
In this case, $V$ and $\tilde{V}$ become
\begin{align}
V &= V_{k} = \frac{2k+n-2}{16}\Big(4(2k+n)\chi^2\lambda^2 
- 4 c_1 (2k+n)\chi \lambda
+c_1^2 (2k+n-2) - 8 c_0 \chi
\Big),
\notag
\\
\tilde{V} &=  \tilde{V}_{k} = \frac{2k + n -2}{16}\Big(
4(2k + n -4)\chi^2\lambda^2 
- 4 c_1 (2k+n -4)\chi \lambda
+c_1^2 (2k+n-2) + 8 c_0 \chi
\Big).
\notag
\end{align}
The quantum-mechanical systems for $H = H_k = A_k^\dag A_k$ 
and $\tilde{H} = \tilde{H}_{k} = A_k A_k^\dag$ 
are
\begin{align}
H_k \psi_{k} &= E_k \psi_{k},
\\
\tilde{H}_{k} \tilde{\psi}_{k} &= \tilde{E}_k \tilde{\psi}_{k},
\label{h2eqno3}
\end{align}
where $E_k$ is
\begin{align}
E_k = \nu^2 + \frac{k(k+n-2)(c_1^2 + 4 c_0 \chi)}{4},
\end{align}
and $\tilde{E}_k$ is a constant.

As shown in Sec.~\ref{sec:susyquantummechanics}, 
$H_{k}$ and $\tilde{H}_{k}$ are connected by the transformation 
$(m^2_k, \delta m^2_k, c_k, \sigma_k) \to (m^2_k + \delta m^2_k, -\delta m^2_k, -c_k, \sigma_k)$.
Since $E_k$ is not changed under this transformation, the KGE with $m^2_k + \delta m^2_k$
becomes Eq.~\eqref{h2eqno3} with $\tilde{E}_k = E_k$.
So, $\tilde{H}_{k}A_k \psi_{k} = E_k A_k \psi_{k}$ implies that 
$A_k$ maps the solution of KGE with $m^2_k$ into that with $m^2_k + \delta m^2_k (= m^2_{k-1})$.\footnote{
We can also see this explicitly as follows. 
The transformation 
$(m^2_k, \delta m^2_k, c_k, \sigma_k) \to (m^2_k + \delta m^2_k, -\delta m^2_k, -c_k, \sigma_k)$
corresponds to $k \to -k-n+2$.
Since we have $H_{-k-n+2} = \tilde{H}_{k}$, we can show
$
H_{-k-n+2} A_k \psi_{k} = \tilde{H}_{k} A_k \psi_{k} 
= A_k H_k  \psi_{k} 
= E_k A_k  \psi_{k} = E_{-k-n+2} A_k  \psi_{k},
$
where we used $ \tilde{H}_{k} A_k = A_k H_k$ and $E_{-k-n+2} = E_k$.
From the relation $m^2_{-k-n+2} = m^2_{k-1}$,
we can see that the solution of the KGE with $m^2_{k}$ is mapped into that with $m^2_{k-1}$ by $A_k$.
}

We can also see that the quantum-mechanical systems have a shape invariance
\begin{align}
\tilde{V}_{k+1} &= V_{k} + \epsilon_k,
\end{align}
with
\begin{align}
\epsilon_k = \frac{(2k + n -1)(c_1^2 + 4 c_0 \chi)}{4},
\end{align}
then we have 
\begin{align}
\tilde{H}_{k+1}  = H_k + \epsilon_k.
\end{align}
Using this, we can show
\begin{align}
\tilde{H}_{k+1} \psi_{k}  = (H_k + \epsilon_k)\psi_{k} = (E_{k} + \epsilon_k) \psi_{k} 
=  E_{k + 1} \psi_{k}.
\end{align}
From 
the relation $A_{k+1}^\dag \tilde{H}_{k+1} = H_{k+1} A_{k+1}^\dag$ and the
above equation, we have
\begin{align}
H_{k+1} A_{k+1}^\dag \psi_{k} 
=  E_{k+1} A_{k+1}^\dag \psi_{k}
\end{align}
This corresponds to the map between solutions of two KGEs with $m^2_{k}$ and $m^2_{k +1}$.

Thus, for a given $k$, we have two ladder operators, 
one shifts $k$ to $k-1$ by the supercharge, 
the other shifts $k$ to $k + 1$ due to the shape invariance.
Repeating this process, we can shift $k$ to $k\mp 1, k\mp 2, \cdots$.
This is the explanation from the point of view of the supersymmetric quantum mechanics 
with a shape invariance potential.

\section{In the presence of a gauge field and a scalar potential}
\label{generalization:gaugefield}
When a maxwell field $A_\mu$, which satisfies $\nabla^\mu F_{\mu \nu} = 0$ with $F_{\mu\nu}=\nabla_\mu A_{\nu} - \nabla_\nu A_\mu$, and a scalar potential $U$ are present, 
the field equation is given by $ ({\cal H }  - m^2) \Phi = 0$, where
the operator ${\cal H }$ is 
\begin{equation}
{\cal H } = g^{\mu\nu}(\nabla_\mu + ie A_\mu)(\nabla_\nu + ie A_\nu) + U \,, \label{Hamiltonian_2}
\end{equation}
and $e$ is an electric charge. 
In analogy with Eq.~\eqref{ladderdk}, we consider the operators
\begin{equation}
D_k := \zeta^\mu(\nabla_\mu + ie A_\mu) - k Q \,, \label{modified_ladder}
\end{equation}
where $\zeta^\mu$ is a CCKV satisfying $\nabla_\mu\zeta_\nu = Q g_{\mu\nu}$, 
and $k$ is a real parameter. We also assume that $\zeta^\mu$ is non-null
and an eigen vector of the Ricci tensor $R_{\mu \nu}\zeta^\nu = (n-1)\chi \zeta_\mu$.
Under these conditions, we discuss the conditions that the operators (\ref{modified_ladder}) become the ladder operators which satisfy the commutation relation \eqref{generalcommutationrelation}.
After a calculation, we obtain
\begin{eqnarray}
[{\cal H },D_k]
&=& \chi(2k+n-2) D_k + 2Q({\cal H } + \chi k(k+n-1)) \nonumber\\
& & + 2 ie \zeta_\nu F^{\mu\nu}(\nabla_\mu + i e A_\mu)
- \zeta^\mu\partial_\mu U -2QU \,. \label{eq10-10}
\end{eqnarray}
Hence, imposing the additional conditions\footnote{
Note that if we impose the condition $F^{\mu\nu} \zeta_\nu = i \sigma \zeta^\mu $ with 
a constant $\sigma$, we can show $\sigma = 0$ by taking an inner product with this condition and $\zeta_\mu$. 
}
\begin{align}
F^{\mu\nu} \zeta_\nu = 0, 
 \qquad \zeta^\mu\partial_\mu U = -2QU,
 \label{condition_ladder_3}
\end{align}
the commutation relation comes down to
\begin{equation}
[{\cal H }, D_k]
= \chi(2k+n-2) D_k + 2Q({\cal H } + \chi k(k+n-1)) \,. \label{com_rel_3}
\end{equation}
This is the commutation relation for the mass ladder operator with 
$m^2 = - \chi k(k+n-1)$ and $\delta m^2 =  \chi(2k+n-2)$.


\begin{thebibliography}{99}
        
\bibitem{Cardoso:2017qmj} 
  V.~Cardoso, T.~Houri and M.~Kimura,
  Phys.\ Rev.\ D {\bf 96}, no. 2, 024044 (2017)
  [arXiv:1706.07339 [hep-th]].


\bibitem{Evnin:2015gma} 
  O.~Evnin and C.~Krishnan,
  Phys.\ Rev.\ D {\bf 91}, no. 12, 126010 (2015)
  [arXiv:1502.03749 [hep-th]].


\bibitem{Evnin:2015wyi} 
  O.~Evnin and R.~Nivesvivat,
  JHEP {\bf 1601}, 151 (2016)
  [arXiv:1512.00349 [hep-th]].


\bibitem{Evnin:2016nsz} 
  O.~Evnin and R.~Nivesvivat,
  J.\ Phys.\ A {\bf 50}, no. 1, 015202 (2017)
  [arXiv:1604.00521 [nlin.SI]].


\bibitem{Bizon:2011gg} 
  P.~Bizon and A.~Rostworowski,
  Phys.\ Rev.\ Lett.\  {\bf 107}, 031102 (2011)
  [arXiv:1104.3702 [gr-qc]].



\bibitem{Aretakis:2011ha} 
  S.~Aretakis,
  Commun.\ Math.\ Phys.\  {\bf 307}, 17 (2011)
  [arXiv:1110.2007 [gr-qc]].


\bibitem{Aretakis:2011hc} 
  S.~Aretakis,
  Annales Henri Poincare {\bf 12}, 1491 (2011)
  [arXiv:1110.2009 [gr-qc]].


\bibitem{Lucietti:2012xr} 
  J.~Lucietti, K.~Murata, H.~S.~Reall and N.~Tanahashi,
  JHEP {\bf 1303}, 035 (2013)
  [arXiv:1212.2557 [gr-qc]].





\bibitem{Higuchi:1986wu}
  A.~Higuchi,
  J.\ Math.\ Phys.\  {\bf 28}, 1553 (1987)
  Erratum: [J.\ Math.\ Phys.\  {\bf 43}, 6385 (2002)].


\bibitem{Gelderen:1998}
M. van Gelderen 1998
DEOS Progress Letter 98.1 (Delft University Press, editied by R. Klees) 57-67.



\bibitem{Batista:2014uja} 
  C.~Batista,
  Class.\ Quant.\ Grav.\  {\bf 31}, 165019 (2014)
  [arXiv:1405.4148 [gr-qc]].


\bibitem{Breitenlohner:1982jf} 
  P.~Breitenlohner and D.~Z.~Freedman,
  Annals Phys.\  {\bf 144}, 249 (1982).


\bibitem{Breitenlohner:1982bm} 
  P.~Breitenlohner and D.~Z.~Freedman,
  Phys.\ Lett.\  {\bf 115B}, 197 (1982).



\bibitem{Wald:1978vm} 
  R.~M.~Wald,
  Phys.\ Rev.\ Lett.\  {\bf 41}, 203 (1978).



\end{thebibliography}
\end{document}